\documentstyle[preprint,aps]{revtex}
\draft
\begin{document}

\title{Effects of N$^{*}$(1440) resonance on subthreshold
kaon, antikaon, and antiproton production}
\bigskip
\author{B. A. Li, C. M. Ko and G. Q. Li}
\address{Cyclotron Institute and Department of Physics,\\
Texas A\&M University, College Station, TX 77843}
\maketitle

\begin{abstract}
We study the effects of $N^{*}(1440)$ resonance on the production of
kaons, antikaons and antiprotons in heavy-ion collisions at subthreshold
energies. Using free-space widths for the $\Delta(1232)$ and the $N^{*}(1440)$
resonance and free-space thresholds for particle production, it is found
that the $N^{*}$-baryon interactions contribute about 10\%, 25\% and 80\%
to the yield of kaon, antikaon and antiproton, respectively. Doubling
the widths of resonances due to possible medium effects, both the
total and the $N^{*}$ induced antiproton yield decrease by about a
factor of 3. Using a reduced antiproton production
threshold in a nuclear medium as a result of dropping
in-medium nucleon and antiproton masses,
the total and the $N^{*}$ induced antiproton yield
increase by a factor of 35 and 27, respectively, and the contribution of
the $N^{*}$ induced collisions is reduced to about 50\% of the total.
\end{abstract}

Particle production, especially at subthreshold energies,
provides valuable information about the high density phase of
relativistic heavy-ion collisions\cite{aich85}.
Experimental data accumulated
at several laboratories during the last decade together with
those obtained more recently at SIS/GSI on the production of
kaons, antikaons, antiprotons and etas as well as pions indicate
that a gradual transition to a resonance matter
occurs in the participant region of heavy-ion collisions at beam energies
of 1 to 2 GeV/nucleon \cite{metag}. The resonance matter is characterized by
a high baryon density with about 1/3 of the nucleons excited to
the resonance
states. To study in detail the properties of this new form of matter, it is
important to first determine its baryonic composition, i.e., the relative
abundance
of nucleons and various baryon resonances. In the beam energy range of
1-2 GeV/nucleon,
the important baryonic resonances are $\Delta(1232)$,
$N^{*}(1440), N^{*}(1520)$ and $N^{*}(1535)$. The excitation of
the $\Delta(1232)$ resonance in heavy-ion collisions
has been extensively studied through pion production, and
its effects on the production of kaons, antikaons as well as
antiprotons at subthreshold energies have been stressed frequently
in the literature. While the excitation of $N^{*}(1535)$ resonance has been
studied through eta mesons \cite{mosel,berg}, no detailed study on the
excitation of $N^{*}(1440)$ resonance
has been carried out, and its effects on subthreshold particle production
are still unclear.

Resonances are produced in heavy-ion collisions from energetic
hadron-hadron collisions.
Subsequent collisions or decays of these high energy resonances
can produce particles that cannot be produced in the
first-chance nucleon-nucleon interactions. From energy consideration, one
expects that the effects of baryon
resonances on particle production strongly depend on the threshold energies
of the particles to be produced,
the mass of resonances, and the excitation functions of
resonances in hadron-hadron collisions.

Also, the properties of hadrons, including
the baryon resonances,  in a hot and dense matter may be different
from those in free space \cite{brown,kaplan,hatsuda,asakawa}.  To study
the properties of hadrons in the hot and dense matter has been one of
the motivations for relativistic heavy-ion research.
It has been shown recently via
the relativistic transport model that the attractive scalar mean-field
potential, which leads to a reduction of particle production thresholds
in a medium, is required
to account for the measured yields of kaons \cite{kok}, antikaons
\cite{kop}, and antiprotons \cite{koa}
from heavy-ion collisions at subthreshold energies.  In these studies,
higher resonances other than the delta have been ignored.  To extract
more quantitatively information of the scalar mean-field potential for
kaons, antikaons, and antiprotons from heavy-ion collisions, it is
important to know the contribution of higher resonances to their production.

In this paper, we shall study the excitation of $N^{*}(1440)$ resonance
in the
hot and dense hadronic matter formed in the initial compression
stage of heavy-ion collisions at
beam energies around 1-2 GeV/nucleon.  In particular, the
effects of $N^{*}(1440)$ resonance on the production of kaons,
antikaons and antiprotons in heavy-ion collisions at subthreshold energies
will be investigated. Our study is based on the hadronic transport model
for heavy-ion collisions, as described in details in refs. \cite{li91a,li91b}.
This model includes both $\Delta(1232)$ and $N^*(1440)$ resonances.
Although the $N^{*}(1520)$ and $N^*(1535)$ resonance are not in the model,
their effect has been partially included as the
nucleon-nucleon inelastic cross section is assumed to be saturated
by the $\Delta(1232)$ and $N^*(1440)$ resonances.

$N^{*}(1440)$ resonances are mainly produced in high energy $\pi N$ and
$NN$ collisions. Using the isospin decomposition of the pion
production cross section in nucleon-nucleon collisions \cite{verwest},
one can see that the $N^{*}(1440)$ resonance starts being excited
significantly at $E^{nn}_{cm}\approx 2.5$ GeV and increases rapidly with
increasing nucleon-nucleon center-of-mass energy.  As the
excitation of the $\Delta(1232)$ resonance decreases with energy in
this energy range, the $N^*$ resonance becomes more important than
the $\Delta$ resonance at $E^{nn}_{cm}\approx 3.0$ GeV.

We first show in Fig.\ 1 the
multiplicity of $N^*(1440)$ resonance from Au+Au collisions
at beam energies between
1 and 2 GeV/nucleon and impact parameters between 1 and 9 fm.
The abundance of $N^{*}(1440)$ resonance strongly depends on the impact
parameter and the beam energy. In central
collisions, the maximum number of $N^{*}(1440)$ resonance increases
from 7 to 17 as the beam energy increases from 1 to 2 GeV/nucleon.
For comparisons, it should be mentioned
that the maximum number of $\Delta(1232)$ resonance increases from 52 to 89
in the same collision, and therefore the ratio of the abundance for the
$N^{*}$ and the $\Delta$ resonance
increases from 14\% to 19\% as the beam energy
increases from 1 to 2 GeV/nucleon.

We now study the effects
of $N^{*}$ resonance on the production of kaons,
antikaons and antiprotons in heavy-ion collisions
at subthreshold energies. The production probabilities of these
particles are calculated perturbatively in the hadronic transport model
as in other model calculations \cite{kok,kop,koa,giessen,huang}.
Since we are only interested in studying the effects of $N^{*}$ resonance,
no final-state interaction for the produced particles is taken into account.
Therefore, the calculated particle production probabilities should be
regarded as primordial ones. Elementary cross sections for the production
of kaons, antikaons and antiprotons are taken as those commonly used in the
literature (e.g. \cite{kok,kop,koa}). It should be noted that in the
calculation for kaons we have assumed that the kaon
production cross sections in $N^{*}$ involved baryon-baryon collisions
are the same as those in $\Delta$ induced collisions. For the production of
antikaons and antiprotons,  we assume as in the literature that the cross
sections of baryon resonance induced collisions are the same as
that of {\it pp} collisions.

Our results on the production of kaons in Au+Au collisions at
beam energies between 1 and 2 GeV/nucleon and impact parameters
between 1 and 9 fm are shown in Fig.\ 2. The
total kaon production probability includes
contributions from $NN, N\Delta, NN^{*}, \Delta\Delta, N^{*}N^{*},
\Delta N^{*}$ and $N\pi$ collisions. In the figure, the total probabilities
are shown with the solid lines while the contributions from the $N^{*}$
involved
collisions are shown with the dashed lines. It is seen that,
the $N^{*}$ involved collisions contribute only about 11\% in
the whole energy and impact parameter ranges. This result is not
surprising since baryon-baryon collisions at
center-of-mass energies around the kaon production threshold of 2.55 GeV
are dominated by the production and absorption of $\Delta$ resonances.
However, by considering the excitation functions of $\Delta$ and
$N^{*}$ resonances in both $\pi N$ and $N N$ collisions one expects
that the $N^{*}(1440)$ resonance would become more effective than
the $\Delta(1232)$ resonance in accumulating enough energy such that
subsequent collisions of $N^*(1440)$ resonances can produce
particles with higher thresholds. This is indeed seen in our
calculations as we will discuss in the following.

Antikaon and antiproton production probabilities in the Ni+Ni
collisions at beam energies of 1.85 GeV/nucleon and 2.1 GeV/nucleon
and impact parameters between 1 and 7 fm are shown in Fig.\ 3 and
Fig.\ 4, respectively.
It is seen that at both beam energies the $N^{*}$ involved collisions
contribute about 25\% to the total antikaon yields.
While for antiproton production the $N^{*}$ involved
collisions contribute about 90\% at 1.85 GeV/nucleon, this
contribution decreases to about 70\% at 2.1 GeV/nucleon as one would expect.
This finding indicates that antiproton production at subthreshold energies
may serve as a possible probe of $N^{*}$ resonances in
the resonance matter formed in relativistic heavy-ion collisions.

However, before deliberating further on the effects of $N^{*}$ resonance
on subthreshold particle production it is necessary to add a few cautions.
In the above calculations, free-space properties of baryon resonances,
i.e. their masses and widths, are used. In nuclei the decreasing mass and
broadening width of $\Delta$ resonance are well known \cite{oset} and have
been a subject of much interest. On the contrary, little is known about
the in-medium behaviour of higher resonances. The very recent
analysis of photoabsorption cross sections on nuclei
for photon energies between 500 and 1500 MeV
has shown that widths of higher resonances in medium are almost
twice as large as that in free space \cite{plb}, and thus there is a large
overlapping between resonances. To see how the broadening of
baryon resonances may change the effects of $N^{*}$ resonance
discussed above, we have performed calculations by doubling the
widths of $\Delta$ and $N^{*}$ resonance.
These calculations can, however, only be regarded as schematic since it is not
reliably known how the widths may depend on the density of the nuclear medium,
and moreover it is highly doubtful that one can still treat resonances
with such large widths as quasiparticles as usually done in transport models.
Results of these calculations for the Ni+Ni collision at a beam energy of
2.1 GeV/nucleon and an impact parameter of 1.0 fm are listed in Table 1.

It is seen that by doubling the widths of the resonances the total
kaon production probability is slightly increased and the contribution
from the $N^{*}$ involved collisions decrease to about 7\%.
This is understandable as the resonances decay faster contributions
to kaon production from higher resonances become less important. The
overall reduction of the kaon production probability in the
resonance involved collisions is, however, almost completely compensated
by the increase of that in the $\pi N$ collisions. For antikaons
the $N^{*}$ contribution decreases from 28\% to 20\%.
For antiprotons, doubling the resonance widths results in a reduction of
both the total and the $N^{*}$ involved production probability by about a
factor of 3. Furthermore, the $N^{*}$ contribution is reduced from 70\%
to 65\%. Nevertheless, we need to be cautious in stressing the
total change of the
antikaon and antiproton yields as contributions to antikaon
and antiproton production from pion induced reactions
are almost unknown and have been neglected in the present calculations.

Another important in-medium effect is the changing masses of hadrons in
a dense medium. This effect on the production of subthreshold kaons,
antikaons, and antiprotons has been studied extensively \cite{kok,kop,koa}.
As the $N^{*}$ resonance is found to have the most important
effect on subthreshold antiproton production, we now study schematically
how this effect may change if in-medium masses of
hadrons are used. For this purpose, we use the empirical
density-dependent nucleon effective mass in a medium as proposed in
refs. \cite{brown,asakawa} for the two produced particles in the
reaction $BB\rightarrow NNp\bar{p}$. Consequently, the
density-dependent threshold for antiproton production in a nuclear
medium can be written as
\[ T_{\bar{p}}=2m_{n}[1+1/(1+0.25\rho/\rho_{0})],
\]
here $m_{n}$ is the free nucleon mass and $\rho_{0}$ is the normal nuclear
matter density. Using the reduced, density-dependent threshold we
have found that the total and the $N^{*}$ involved antiproton
yield increase by a factor of 35 and 27, respectively,  in
the Ni+Ni collision at a beam energy of 2.1 GeV/nucleon and an impact
parameter of 1.0 fm. The contribution of the $N^{*}$ involved
collisions decreases to about 50\% of the total probability.
Numerical results of this calculation are listed in the last row
of Table 1.

In summary, we have studied the excitation of the $N^{*}(1440)$
resonance in heavy-ion collisions at beam energies from 1 to 2 GeV/nucleon.
It is found that the contribution of $N^{*}$ resonance to subthreshold kaon
and antikaon production is small compared with that from the delta
resonance.  The results of Refs. \cite{kok,koa} obtained without the
$N^*$ resonance are therefore not much affected.
On the other hand, the $N^*$ resonance does contribute appreciably
to antiproton production from heavy-ion collisions at subthreshold energies.
Since the medium effects due to the attractive scalar mean field are
much larger than the effect from the $N^*$ resonance, the conclusion
of Ref. \cite{kop} that the observed antiproton yield in subthreshold
heavy-ion collisions is consistent with a reduced threshold as a result
of the reduction in the in-medium antiproton mass remains valid.
However, to study more quantitatively the in-medium effects on antiproton
production, one needs to include in the future the $N^*$ resonance.
We therefore conclude that
antiproton production from heavy-ion collisions at subthreshold energies
not only provides information on the in-medium properties of
the antiproton but also serve as a possible probe of the $N^{*}(1440)$
resonance in the resonance matter.

\medskip

This research was supported in part by the NSF Grant No. PHY-9212209
and the Welch Foundation Grant No. A-1110.

\begin{table}
\caption{Effects of doubling the resonance widths and reducing the
$\bar{p}$ production threshold}
\begin{tabular}{ccccccccc}
\hline
\multicolumn{1}{c}{Calculations}
&\multicolumn{1}{c}{$P_{K}^{total}$}
&\multicolumn{1}{c}{$P_{K}^{N^{*}}$}
&\multicolumn{1}{c}{$P_{\bar{K}}^{total}$}
&\multicolumn{1}{c}{$P_{\bar{K}}^{N^{*}}$}
&\multicolumn{1}{c}{$P_{\bar{p}}^{total}$}
&\multicolumn{1}{c}{$P_{\bar{p}}^{N^{*}}$}
%$\;\;\;\;\;\;\;\;$\\
\\
\hline
\multicolumn{1}{l}{free widths}
&\multicolumn{1}{c}{0.180}
&\multicolumn{1}{c}{0.020}
&\multicolumn{1}{c}{0.730$*10^{-2}$}
&\multicolumn{1}{c}{0.208$*10^{-2}$}
&\multicolumn{1}{c}{0.827$*10^{-5}$}
&\multicolumn{1}{c}{0.576$*10^{-5}$}
\\
\multicolumn{1}{l}{double widths}
&\multicolumn{1}{c}{0.184}
&\multicolumn{1}{c}{0.013}
&\multicolumn{1}{c}{0.575$*10^{-2}$}
&\multicolumn{1}{c}{0.117$*10^{-2}$}
&\multicolumn{1}{c}{0.295$*10^{-5}$}
&\multicolumn{1}{c}{0.190$*10^{-5}$}
\\
\multicolumn{1}{l}{reduced $\bar{p}$ threshold}
&\multicolumn{1}{c}{}
&\multicolumn{1}{c}{}
&\multicolumn{1}{c}{}
&\multicolumn{1}{c}{}
&\multicolumn{1}{c}{0.346$*10^{-3}$}
&\multicolumn{1}{c}{0.174$*10^{-3}$}
\\
\hline
\end{tabular}
\end{table}

\begin{figure}

\caption{
Time evolution of the $N^{*}(1440)$ population in Au+Au collisions.
(left) Impact parameter dependence. (right) Beam energy dependence.}
\end{figure}
\begin{figure}
\caption{
Total (solid) and $N^{*}(1440)$ induced (dashed) kaon production
probability in the Au+Au collisions. (left) Impact parameter dependence.
(right) Beam energy dependence.}
\end{figure}
\begin{figure}
\caption{
Total (solid) and $N^{*}(1440)$ induced (dashed) antikaon production
probability in Ni+Ni collisions at 1.85 GeV and 2.1 GeV/nucleon.}
\end{figure}
\begin{figure}
\caption{
Total (solid) and $N^{*}(1440)$ induced (dashed) antiproton production
probability in Ni+Ni collisions at 1.85 GeV and 2.1 GeV/nucleon.}
\end{figure}

\end{document}